\documentclass[doublecol]{epl2} 

\usepackage{amsmath}

\title{Tuning attraction and repulsion between active particles through persistence}

\author{M. J. Metson \and M. R. Evans \and R. A. Blythe}

\institute{                    
  SUPA, School of Physics and Astronomy, University of Edinburgh, Edinburgh EH9 3FD, UK
}

\abstract{
We consider the interplay between persistent motion, which is a generic property of active particles, and a recoil interaction which causes particles to jump apart on contact.
The recoil interaction exemplifies an active contact interaction between particles, which is inelastic and is generated by the active nature of the constituents. It is inspired by the `shock' dynamics of certain microorganisms, such as \emph{Pyramimonas octopus}, and always generates an effective repulsion between a pair of passive particles. Highly persistent particles can be attractive or repulsive, according to the shape of the recoil distribution. We show that the repulsive case admits an unexpected transition to attraction at intermediate persistence lengths, that originates in the advective effects of persistence. This allows active particles to fundamentally change the collective effect of active interactions amongst them, by varying their persistence length.}

\begin{document}

\maketitle

{\em Introduction} -- Active matter is composed of particles that consume energy to move with a persistent velocity \cite{Marchetti2013}. Examples include birds \cite{Cavagna2010}, fish \cite{Parrish2002}, self-phoretic particles \cite{Paxton2004} and swimming microorganisms \cite{Schnitzer1993}. The internal driving that allows the direction of motion to persist, combined with steric or hydrodynamic interactions between particles, can generate distinctively nonequilibrium states of matter \cite{Viksek2012,Marchetti2013,Cates2015,Elgeti2015,Julicher2018}. Notably, when a particle's motion is blocked by short-range repulsion from obstacles, persistence engenders an effective attraction \cite{Slowman2016,Slowman2017} which in turn can precipitate a motility-induced phase separation \cite{Cates2015}. In the widely-studied paradigms of run-and-tumble and active Brownian particles, it is usually found that increasing persistence generates more blocking and therewith an ever-stronger attraction \cite{Solon2015}.

Theoretical models of active particles typically assume a passive contact interaction, for example, one that derives from a potential with a short-range repulsion through reversible energy exchange with a heat bath \cite{Fily2012,Redner2013}. Much less explored are \emph{active} contact interactions, that is, those that arise from internal driving. In this work, we determine the effects of persistence in the presence of an active recoil contact dynamics, a process that is inspired by the `shock’ gait exhibited by microorganism \emph{Pyramimonas octopus} \cite{Wan2016,Wan2018}. This process is characterized by an extremely rapid movement, which here we take to be triggered by a collision between two particles.
This interaction is inherently active in that it requires expenditure of energy for the act of recoil and is thus strongly inelastic. We show that this
active interaction generates physical phenomena that are fundamentally distinct from those described above for passive contact interactions. First, we establish that actively repulsive particles are able to retain their repulsive character in the face of arbitrarily high persistence, in contrast to short-range repulsive particles. Second we find, despite an expected effective repulsion at low persistence, that even a strongly-repulsive recoil can be transformed into an attraction at intermediate persistence lengths. These results imply that a tunable persistence length for an active particle could be exploited  to engineer on demand an effective attraction to, or repulsion from, other objects to achieve some goal.

We establish the possibilities created by recoil and persistence by solving exactly for the stationary interparticle distribution function for a pair of persistent random walkers in a periodic one-dimensional system, from which an effective attraction or repulsion between them can be identified. Despite persistence being a core property of active particles, exact results for \emph{interacting} particles at the microscopic scale are very sparse 
\cite{Slowman2016,Slowman2017,Mallmin2019,Das2020,LeDoussal2021,Singh2021a} compared to the noninteracting case (e.g.~\cite{Angelani2014,Malakar2018,Demaerel2018,Hartmann2020,Malakar2020,Mori2020,Mori2021,DeBruyne2021,Singh2021b}). Enlarging the set of exact results for interacting systems constitutes a key step towards a physical theory for many-body systems of active matter that is fully grounded in microscopic principles and eludes us at present \cite{Cates2022}.

\begin{figure}[tb]
\begin{center}
\includegraphics[width=0.75\linewidth]{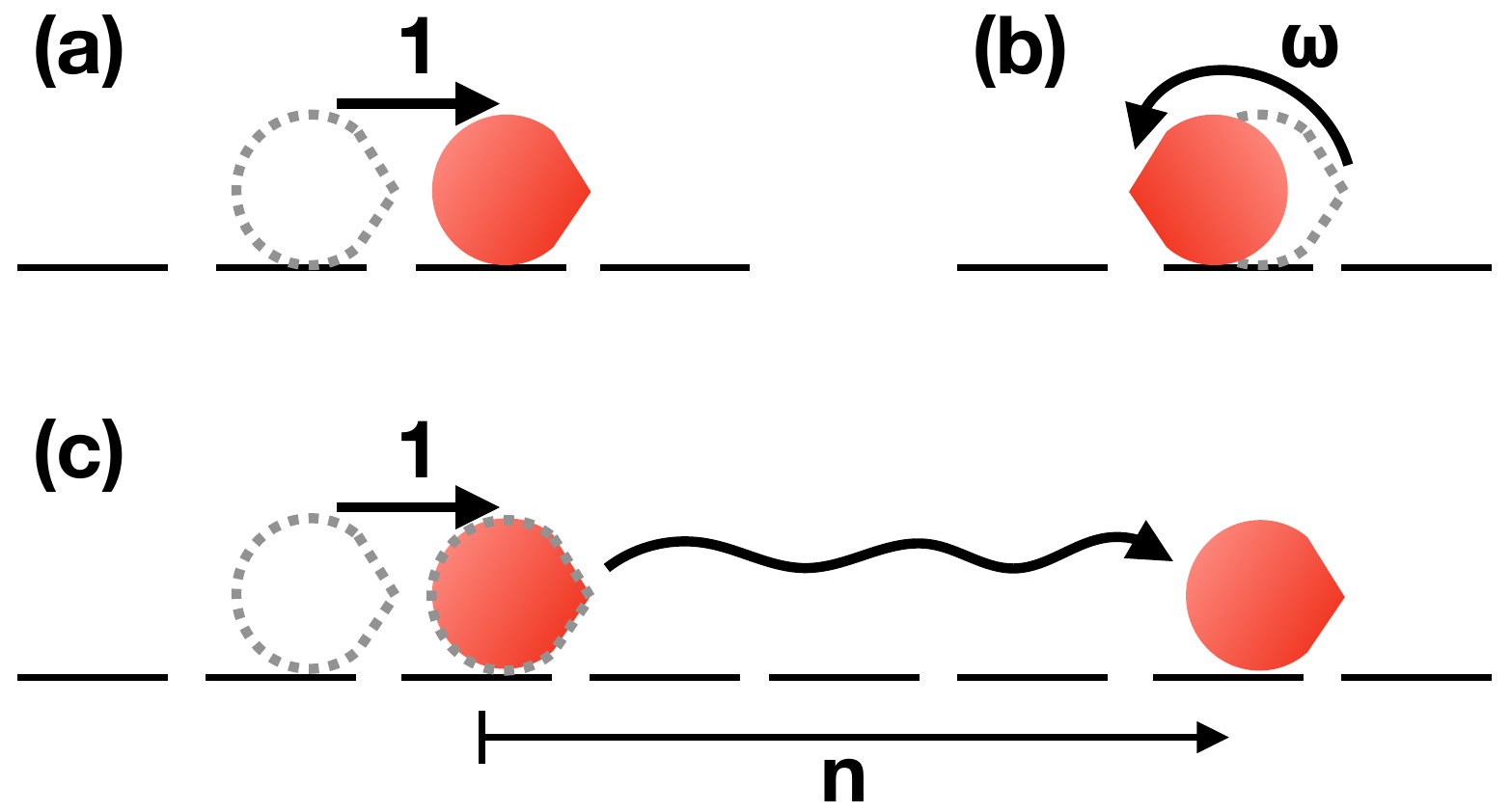}
\end{center}
\caption{\label{fig:model} Model dynamics. Dashed outlines indicate the configuration before a transition, and the pointed end the direction of motion. (a) A hop in the current direction at rate $1$. (b) Spontaneous reversal at rate $\omega$. (c) A hop onto a neighbour, causing the latter to recoil by $n$ sites drawn from the distribution $\Phi(n)$, and then reverse with probability $r$ (not shown).}
\end{figure}

{\em Model definition} -- The model system we solve is defined as follows (see also Fig.~\ref{fig:model}). Two particles reside on a periodic one-dimensional lattice of $L$ sites and each has a direction of motion ($+$ or $-$). Each particle hops to the adjacent site in its direction of motion as a Poisson process with rate $\gamma$, which we shall take to be unity to set the overall timescale, and can spontaneously reorient (i.e., reverse its direction) at rate $\omega$. If a particle hops on top of the other, the latter is displaced by $1\le n < L$ sites in the direction of the hopping particle's motion, where $n$ is drawn from a distribution $\Phi(n)$. For simplicity, the recoil process is chosen to be instantaneous. On landing, the displaced particle reverses its velocity with probability $r$ (thereby retaining it with probability $1-r$). This reversal could arise, for example, from the particle rotating when it recoils. Prominent special cases include $r=0$, where recoil does not induce a reversal and $r=\frac{1}{2}$, where the velocity is randomized on recoil. We find below that certain physical effects emerge only when the reversal rate is nonzero.
We emphasize that the model thus defined entails active interactions, as discussed in the introduction.

{\em Stationary state in a scaling limit} --
In this work we report on the exact stationary state of the model and the novel features it reveals. By stationary state we refer to the long time limit of the probability distributions of the interparticle separations; in the long-time limit these distributions become stationary.  The stationary state is inherently nonequilibrium as detailed balance does not hold, thus there is a circulation of probability between the sector of configuration space in which the particles are following each other and the sector in which they are approaching each other.

Although it is possible to solve for the stationary interparticle distribution functions on the lattice, we obtain the cleanest results in a \emph{scaling limit} where particles move ballistically but reorientations remain stochastic. Formally, this is achieved by taking the number of lattice sites $L\to\infty$ and the reversal rate $\omega\to0$ such that the \emph{persistence length} of the walkers,
\begin{equation}
\xi = \lim_{\substack{L\to\infty\\ \omega \to 0}} \frac{\gamma}{\omega L} \;,
\end{equation}
remains finite. Here, $\xi$ corresponds to the mean fraction of the system traveled by a particle between two reversals (assuming that no collision intervenes). For the recoil distribution $\Phi(n)$, we choose functions for which the limit
\begin{equation}
\rho(x) = \lim_{L\to\infty} L\Phi(Lx)
\end{equation}
is normalized to have  unit integral over the interval $0\le x \le 1$.

We now present the solution for the stationary state in this scaling limit. We distinguish between configurations in which the two particles are following each other, and those in which one is approaching the other. We denote the stationary (time-independent) inter-particle distribution functions that apply in these two sectors as $p(x)$ and $q(x)$, respectively. Note that, by symmetry, $q(1-x)$ gives the corresponding distribution for particles that are a distance $x$ apart and receding. We find for following particles that, up to a normalization, $Z$, 
\begin{equation}
\label{pofq}
p(x) = \frac{1}{2 Z} \left[ q(x)+q(1-x) + r\xi \rho_+(x) \right]
\end{equation}
and for approaching particles
\begin{multline}
\label{q}
q(x) = \frac{1-r}{2Z} \int_x^1 {\rm d} y \rho_-(y) + \frac{1}{2Z}\int_x^1 {\rm d} y \rho_+(y) + \\\frac{1}{2\xi Z} \left[ (1-x) \int_0^x {\rm d} y y\rho_+(y) + x \int_x^1 {\rm d} y (1-y) \rho_+(y) \right]
\end{multline}
where $\rho_{\pm}(x)$ are the (anti-)symmetrized forms of the recoil distribution: $\rho_{\pm}(x) = \rho(x) \pm \rho(1-x)$. These two expressions (\ref{pofq}) and (\ref{q}) are the main result of this work, and apply if the recoil distribution $\rho(x)$ is differentiable at the boundaries. The more general case, and the technical subtleties that arise at the boundaries, is discussed in detail in a forthcoming paper \cite{longshock}.

\emph{Physics of different persistence regimes} --
A notable feature of the distribution functions $p(x)$ (\ref{pofq}) and $q(x)$ (\ref{q}) is that there are contributions at three distinct orders in the persistence length $\xi$.  We consider first the {\em strong-persistence limit}, $\xi\to\infty$. Then, as long as there is some possibility for a particle to reverse on recoil, the last term in (\ref{pofq}) dominates, and there is an `imprint' of the symmetrized recoil distribution, $\rho_+(x)$ in the inter-particle distribution for following particles. The physical origin of this imprint is first that a separation $x$ is entered when approaching particles meet and the displaced particle reverses direction (probability $r$). This fixed separation is maintained for a time proportional to $\xi$ before one of the particles reorients, hence the factor of $\xi$ in the r.h.s.\ of \eqref{pofq}. The recoil distribution is symmetrized due to the invariance under particle relabeling. Importantly, this symmetrized recoil distribution can be concentrated near the boundaries, thus constituting an attraction, or peaked around the mid-point, thus constituting a repulsion. We note that this latter repulsion is not normally seen at high persistence in active matter systems with steric (passive) interactions.

We now turn to the {\em vanishing persistence length limit}, $\xi\to0$, in which particle motion becomes diffusive. Due to the high frequency of reorientations, both distributions approach a limit given by the final line in (\ref{q}). This expression is equivalent to the stationary distribution of a diffusive resetting process \cite{Evans2020} where the position $x$ of the diffusive particle resets to a value drawn from the symmetrized distribution $\rho_{+}(x)$ whenever either of the boundary points $x=0$ or $x=1$ is reached. More precisely, the final term in (\ref{q}) solves the equation $q''(x) = -\rho_{+}(x)$ with absorbing boundaries at $x=0,1$.
The equation $q''(x) -\rho_{+}(x)$ and the boundary conditions,  $q(0)= q(1)=0$, imply the inter-particle distribution functions are both concave and peaked around the midpoint $x=\frac{1}{2}$ due to the non-negativity and symmetry of $\rho_{+}$, respectively. An effective repulsion between the particles is therefore inescapable in the diffusive limit. When the imprint $\rho_{+}$ favors attraction, we thus see that there must be a transition from an effective repulsion to effective attraction as the persistence length increases.

Finally we discuss the {\em intermediate persistence length regime}, $\xi =O(1)$. The first two terms in (\ref{q}) enter at order $\xi^0$. Since the distribution $q(x)$ applies to particles that are moving towards each other, we can understand the integrals over the recoil distribution at separations $y>x$ as originating from the advective effects of persistence. These terms allow for an effective attraction between the particles at intermediate persistence lengths even when the persistent imprint and the diffusive contribution are both repulsive.

{\em Examples of specific recoil distributions} --
Having presented our general results, we now demonstrate the transitions between effective repulsive and attractive interactions with specific forms of the recoil distribution $\rho(x)$. We consider first the exponential distribution $\rho(x) \propto {\rm e}^{-x/\ell}$, which is characterized by the single lengthscale $\ell$. Note that $\ell$ can be negative, which corresponds to longer jumps being more likely than shorter ones. For any $\ell$, the symmetrized distribution $\rho_{+}(x) \propto \cosh(\frac{1-2x}{2\ell})$ is always peaked at the boundaries, and from the foregoing we expect a transition from effective repulsion to attraction above some critical persistence length.

To see this, we consider the marginal inter-particle distribution $Q(x)$ that is obtained by summing the joint distribution functions over all possible velocity configurations at fixed separation $x$: 
\begin{equation}
Q(x) = p(x)+p(1-x) + q(x) + q(1-x)\;.
\end{equation}
Substituting
\begin{equation}
\rho_{+}(x) = \frac{1}{\ell}\frac{\cosh(\frac{1-2x}{2\ell})}{\sinh (\frac{1}{2\ell})}\;\; ; \;\; \rho_{-}(x) = \frac{1}{\ell}\frac{\sinh(\frac{1-2x}{2\ell})}{\sinh (\frac{1}{2\ell})}
\end{equation}
in \eqref{q} and then \eqref{pofq}, and performing the elementary integrals, yields
\begin{align}
Q(x)&\propto Q_0 + \left( r \frac{\xi}{\ell} + 2(1-r) - 2 \frac{\ell}{\xi}\right) \frac{\cosh(\frac{1-2x}{2\ell})}{\sinh(\frac{1}{2\ell})}
\end{align}
in which, for clarity, we have dropped the overall normalization and subsumed a combination of model parameters into the constant term $Q_0$ as this has no impact on whether the effective interaction is attractive or repulsive. When the prefactor of the $\cosh$ function is negative, $Q(x)$ is peaked at $x=\frac{1}{2}$ and the effective interaction is repulsive. When $\ell$ is positive, the criterion for such repulsion to occur is
\begin{equation}
\frac{\xi}{\ell} < \frac{\sqrt{1+r^2} - (1-r)}{r} \;.
\end{equation}
For all values of the reversal probability, including $r=0$, we find that the attractive imprint grows in size with increasing persistence until its attractive character is reflected in the inter-particle distribution function.

When $\ell$ is negative, we require instead that
\begin{equation}
\frac{\xi}{|\ell|} < \frac{\sqrt{1+r^2} + 1-r}{r} 
\end{equation}
if repulsion is to occur. Here, there is a transition from a repulsive to attractive interaction at a finite persistence length if the velocity reversal probability $r>0$. The critical value of the persistence length decreases as $r$ increases.

When $r>0$, a single transition to attraction at high persistence is generic to any recoil distribution $\rho(x)$ that has a positive curvature at the midpoint, i.e., $\rho''(\frac{1}{2})>0$. This can be seen from the second derivative of the marginal inter-particle distribution function
\begin{equation}
\label{Q''}
Q''({\textstyle\frac{1}{2}}) = 2 r\xi \rho''({\textstyle\frac{1}{2}}) - 4(1-r) \rho'({\textstyle\frac{1}{2}}) - \frac{4}{\xi} \rho({\textstyle\frac{1}{2}}) \;.
\end{equation}
Since the first term on the r.h.s.\ is positive and the last negative, this midpoint curvature increases monotonically from a negative value at small $\xi$ (i.e., a repulsive peak) to a positive value at large $\xi$ (an attractive trough). For the special case $r=0$, a transition occurs only when the recoil distribution has a negative gradient at the mid-point, as was the case when $\ell>0$ above.

Less expected is a reentrant transition that can occur when the recoil distribution $\rho_+(x)$ is peaked at the midpoint and leaves a repulsive imprint. Then, the first term in (\ref{Q''}) is negative, and there are two positive values of $\xi$ at which $Q''(\frac{1}{2})=0$ when both $\rho'({\textstyle\frac{1}{2}})<0$ and the inequality
\begin{equation}
\label{reent}
- \frac{(1-r)^2 (\rho'({\textstyle\frac{1}{2}}))^2}{2r\rho({\textstyle\frac{1}{2}})} < \rho''({\textstyle\frac{1}{2}}) < 0
\end{equation}
is satisfied. Between these two values of $\xi$, the inter-particle distribution function $Q(x)$ has a trough at $x=\frac{1}{2}$, even though it is peaked in both the diffusive and highly persistent limits. This attraction at intermediate persistence lengths can be understood in physical terms via the particle separation being advected away from the central region to the boundaries, but not so far that the repulsive imprint dominates.

\begin{figure}[tb]
\begin{center}
\includegraphics[width=0.8\linewidth]{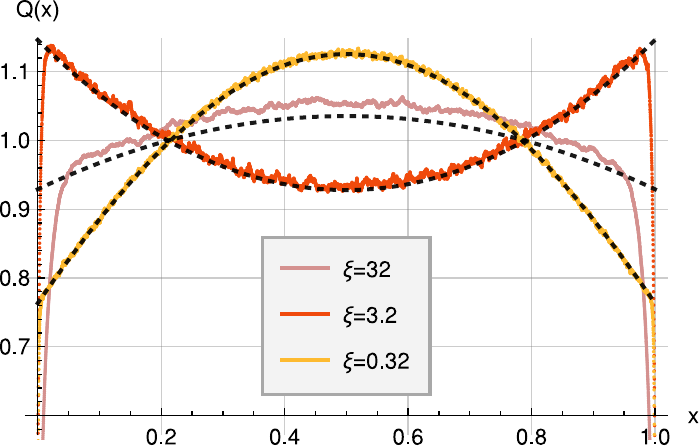}
\end{center}
\caption{\label{fig:reent} Attraction at intermediate persistence lengths $0.825 < \xi < 18.2$ when $\rho(x) = \frac{3}{2}(1-x^2)$, $r=\frac{1}{20}$ and the imprint is repulsive. Analytical predictions (black dashed lines) are compared with distributions obtained from Monte Carlo simulations on a lattice of $L=10^5$ sites and averaged over up to $10^{13}$ units of time. The discrepancy evident for $\xi=32$ is due to finite-$L$ corrections.}
\end{figure}

An example of a recoil distribution that satisfies (\ref{reent}) is the quadratic $\rho(x) = \frac{3}{2}(1-x^2)$. Whilst the symmetrized distribution is peaked at $x=\frac{1}{2}$, one finds an effective attraction over an intermediate range of $\xi$ when $0 < r < \frac{5-\sqrt{21}}{2}\approx0.209$. Fig.~\ref{fig:reent} illustrates for  $r=\frac{1}{20}$ the transition from repulsion to attraction above a lower critical persistence length, and the reversion to a repulsive interaction above an upper critical value.

{\em Derivation of stationary distributions \eqref{pofq},\eqref{q}} --
Having illustrated the nontrivial physical consequences of the distributions (\ref{pofq}) and (\ref{q}), we now outline how they were derived. We return to the lattice model of Fig.~\ref{fig:model} and the master equations for the probability $P_{\sigma_1\sigma_2}(n)$ that the particles have the direction of motion $\sigma_{1,2} = \pm$ and the separation between particle $1$ and $2$ is $n$. For a pair of right-moving particles, $\sigma_{1,2} = +$, we have
\begin{multline}
\label{me1}
\dot{P}_{++}(n) = [P_{++}(n-1)-2P_{++}(n)+P_{++}(n+1)] + {} \\
\shoveleft\quad \omega[ P_{-+}(n) + P_{+-}(n) - 2 P_{++}(n) ] + {} \\
\shoveleft\quad r [ P_{+-}(1) \Phi(n) + P_{-+}(L-1) \Phi(L-n) ] + {} \\
\shoveleft\quad (1-r) [ P_{++}(1) \Phi(n) + P_{++}(L-1) \Phi(L-n) ] \;,
\end{multline}
where $\dot{P}_{++}(n)$ is the time derivative of the probability that the two particles are both right-moving and have separation $n$.
The first line of the r.h.s.\ of this equation comprises contributions from particles moving further apart or closer together, as the particle in front or behind hops with rate $\gamma =1$, respectively. The second line comprises contributions from spontaneous reorientations with rate $\omega$, 
for example $\omega P_{-+}(n)$ is the rate of reorientation of the left particle multiplied by the probability that particles have separation $n$ and direction of motions  $\sigma_1= -$, $\sigma_2 =+$.
The last two lines comprise contributions from recoil 
to separation $n$ from states in which the particles are adjacent ($n=1$ or $n=L-1$). The third line comprises the contributions where the recoiling particle reorientates (with probability $r$) and the fourth line
the contributions where the recoiling particle does not reorientate (probability $1-r$).
The corresponding equation for a pair of left-moving particles is obtained by replacing $P_{++}(n)$ with $P_{--}(n)$ everywhere. This pair of equations holds for all $1<n<L$ if we impose the boundary conditions  $P_{\sigma\sigma}(0)=P_{\sigma\sigma}(L)=0$. Similar considerations yield the master equation for approaching particles
\begin{multline}
\label{me2}
\dot{P}_{+-}(n) = 2[P_{+-}(n+1) - P_{+-}(n)] + \\
\shoveleft\quad \omega[ P_{++}(n) + P_{--}(n) - 2P_{+-}(n) ] + {} \\
\shoveleft\quad \{ r [P_{++}(1) + P_{--}(1)] + 2(1-r) P_{+-}(1) \} \Phi(n) \;,
\end{multline}
with that for receding particles obtained via the replacement $P_{\sigma_1\sigma_2}(n)$ with $P_{\sigma_2\sigma_1}(L-n)$. The boundary conditions on these equations are $P_{+-}(L)=P_{-+}(0)=0$.

In the scaling limit, the inter-particle distribution functions are
\begin{equation}
\label{pq}
p(x) = \lim_{L\to\infty} \frac{P_{++}(Lx)}{P_{+-}(1)} \quad\mbox{and}\quad
q(x) = \lim_{L\to\infty} \frac{P_{+-}(Lx)}{P_{+-}(1)}
\end{equation}
where the right-hand sides are given by the stationary solution of (\ref{me1}) and (\ref{me2}). At stationarity, there are two important symmetries. First, invariance under particle relabelling implies $P_{\sigma_1\sigma_2}(n)=P_{\sigma_2\sigma_1}(L-n)$. Thus $p(x)$ is a symmetric function $p(x)=p(1-x)$, and  the limiting form of the distribution for receding particles ($-+$) is given by $q(1-x)$. Second, parity invariance implies $P_{++}(n)=P_{--}(L-n)=P_{--}(n)$, and so the limiting form of the $--$ distribution is $p(x)$. In other words, the two densities $p(x)$ and $q(x)$ contain all information about the stationary distribution.

Substituting for $p$ and $q$ in (\ref{me1}), and expanding in powers of $\frac{1}{L}$, we find that the stationary distribution for particles that are following each other satisfies
\begin{equation}
\label{sme1}
\frac{1}{L} p''(x) + \frac{1}{\xi} [ q(x)+q(1-x)-2p(x)] = - [ r + (1-r)\kappa_L ] \rho_+(x)
\end{equation}
in which $\kappa_L$ is defined as $\kappa_L = \frac{P_{++}(1)}{P_{+-}(1)}$ for a system size $L$. From (\ref{me2}) we find
\begin{equation}
\label{sme2}
q'(x) + \frac{1}{\xi} [p(x)-q(x)] = - [r \kappa_L + (1-r)] \rho(x) \;.
\end{equation}
In a forthcoming work \cite{longshock}, we solve a system equivalent to (\ref{sme1}) and (\ref{sme2}) for a general recoil distribution $\rho(x)$, including those that have delta functions at the boundary points $x=0$ and $x=1$, which correspond to the particles having some finite probability of jamming into, or passing through, each other on contact. This solution has complex behavior in the region of size $\frac{1}{\sqrt{L}}$ at each boundary, with discontinuities in the interparticle distribution functions and/or their derivatives at the boundary points. When the recoil distribution $\rho(x)$ is differentiable at the boundaries, we find that $\kappa_L\to0$ as $L\to\infty$, and outside the boundary region, the second derivative term in (\ref{sme1}) can be dropped. Rearranging  (\ref{sme1}) then delivers our first main result, Eq.~(\ref{pofq}).

To find $q(x)$, we combine (\ref{sme2}) and (\ref{pofq}) to obtain
\begin{align}
q'(x)+q'(1-x) &= -\rho_+(x) \label{qq'+} \\
q'(x)-q'(1-x) - \frac{1}{\xi} [q(x)-q(1-x)] &= - (1-r)\rho_-(x) \label{qq'-} 
\end{align}
A further simplification that occurs when $\rho(x)$ is differentiable at the boundaries is that the boundary conditions $q(0)=1$ and $q(1)=0$ carry over directly from the master equation (\ref{sme2}). The fact that this does not occur in the general case is addressed in depth in \cite{longshock}. Integrating (\ref{qq'+}) with these boundary conditions gives
\begin{equation}
\label{qdiff}
q(x) - q(1-x) = 1 - \int_0^x {\rm d}y \rho_+(y) \;.
\end{equation}
Substituting this into (\ref{qq'-}) and integrating again leads to the second main result, Eq.~(\ref{q}).

Finally, it is worth briefly discussing the normalization, $Z$, appearing in (\ref{pofq}) and (\ref{q}). Integrating the latter equation we obtain a constant
\begin{equation}
Z_q = \int_0^1{\rm d}x q(x) = (1-r)\bar{y} + \frac{r}{2} + \frac{1}{\xi}\overline{y(1-y)} \end{equation}
where an overline denotes an average with respect to the recoil distribution $\rho(x)$ (an integral over $x \in [0,1]$ weighted by $\rho(x)$). Then, the corresponding integral over $p(x)$ is
\begin{equation}
Z_p = \int_0^1{\rm d}x p(x) = Z_q + r\xi \;.
\end{equation}
Using $Z=Z_p$ in (\ref{pofq}) and $Z=Z_q$ in (\ref{q})  then gives the distribution of separation \emph{conditioned on} being in the following or approaching state. Alternatively, we use $Z=2(Z_p+Z_q)$ for both $p(x)$ and $q(x)$  to obtain the \emph{joint} distribution of being in a specific velocity configuration (e.g., both particles moving to the right) and at a separation $x$.

\begin{figure}[tb]
\begin{center}
\includegraphics[scale=0.55]{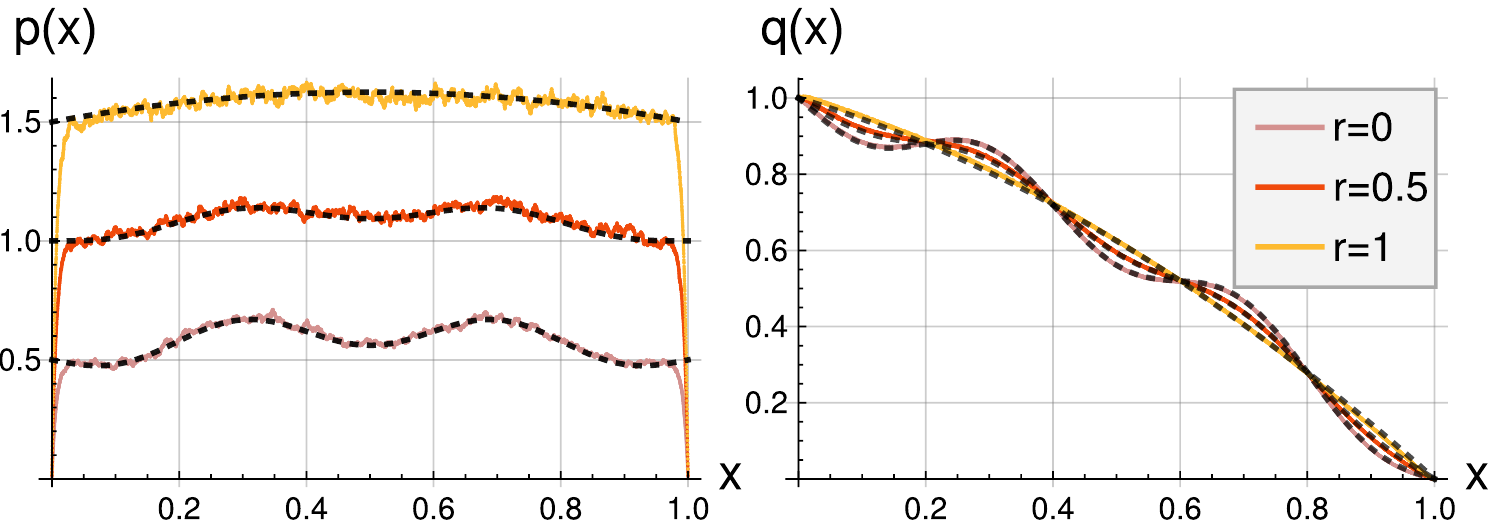}
\end{center}
\caption{\label{fig:pqsim} Comparison between analytical (dashed black lines) and numerical $p(x)$ and $q(x)$ for $\rho(x)=\cos{(5\pi x)}+1$, $\xi=1$, $L=10^4$ and various $r$. Monte Carlo distributions were obtained by averaging over $10^{10}$ units of time.}
\end{figure}

We confirm the validity of (\ref{pofq}) and (\ref{q}) in Fig.~\ref{fig:pqsim} by choosing a form of the recoil distribution with multiple peaks and troughs and comparing with simulations of the lattice-based process. We find excellent agreement, except in the boundary regions where the discarded second derivative in (\ref{sme1}) becomes important.

{\em Summary and conclusion} --
To summarize, in this work we have investigated the inter-particle distribution function that emerges in a system of persistent random walkers that recoil on contact to a distance drawn from the distribution $\rho(x)$. Our main finding is that while the effective interaction that emerges is generically repulsive---and always so when the persistence length is sufficiently short---there are conditions under which an attractive interaction can be engineered by tuning the persistence length. By repulsive and attractive here we mean that the inter-particle distribution function has a maximum or a minimum at the midway point, respectively.

For recoil distributions with a positive curvature at the midpoint, such as an exponentially decaying distribution, there is a transition from repulsion to attraction above a critical persistence length, \emph{similar to what is found for particles with steric contact interactions}. This transition can be understood in terms of a competition between a diffusive resetting process (which generates an effective repulsion) and the persistent imprint of the recoil distribution (which is attractive when the curvature is positive at the midpoint). Recoil interacts nontrivially with persistence when the imprint of the recoil distribution is repulsive. Then, an effective attraction can emerge at intermediate persistence lengths, as long as the recoil distribution has negative gradient at the midpoint and its curvature satisfies (\ref{reent}). This is a consequence of the advective nature of persistence: if particles move sufficiently quickly away from the post-recoil separation, probability can accumulate at small separations. However, this effect is curtailed as the persistence length is further increased, as this extends the lifetime of the repulsive imprint.

The simulation data in Figs.~\ref{fig:reent} and \ref{fig:pqsim} demonstrate the existence of finite-size corrections at the boundaries that are absent from the solutions (\ref{pofq}) and (\ref{q}), which were obtained for $L\to\infty$. It transpires that these finite-size corrections are in general highly nontrivial. In particular, we find that $p(x)$ is generically discontinuous at the boundaries, as is the derivative of $q(x)$. The upshot of this is that the boundary conditions that naturally arise by considering probability fluxes at the boundaries are not satisfied by the limiting forms of $p(x)$ and $q(x)$. This has implications for constructing continuum descriptions of interacting active matter systems which are explored more fully in a forthcoming paper \cite{longshock}.

In conclusion, our results show  that varying the persistence length of 
active matter constituents can change the nature of an active interaction from repulsive to attractive and back again.  It would be interesting to investigate such possibilities experimentally. 
It is tempting to
speculate that an organism equipped with the ability to tune its persistence length could do so productively to approach or avoid objects in its environment. However, our simplified model,  comprising only two particles in one dimension, is rather far from biological reality.
We now discuss these limitations.

First we mention that in \cite{Metson2020} it was shown how results for two persistent random walkers with hard-core interactions, could be used to approximate a many-particle system in arbitrary spatial dimension and predict the formation of jammed clusters, which have been studied numerically \cite{Soto2014}. It would be an interesting challenge to try to extend this approach to the case of recoil interactions.
Although we have restricted our discussion to the two-particle case in this work, it is possible that the system size could serve as a proxy for the typical inter-particle distance in a many-body system, and it would be worthwhile also to determine theoretically how the persistence length, recoil length and particle density interact. As an example, for parameter choices that give rise to repulsion in the two-particle case, it's possible that a state with uniform spatial distribution may emerge in the corresponding many-body system. Conversely, for attractive parameter choices, one might see clustering instead. 

We have also limited the discussion to a periodic system. It would also be of interest to study the case where particles are confined to a region, for example by hard walls. In the case of persistent particles without recoil, it has been found that particles tend to cluster near the walls \cite{Deblais2018,Sartori2018}. We therefore anticipate that recoil combined with persistence will lead to a similarly nontrivial interaction with stationary obstacles such as walls.

Whilst an analytical description of the many-body system is currently beyond what current techniques can achieve, extensive simulations could help explore such possibilities and investigate other pertinent questions surrounding the many-body system. This we suggest could form the basis for future work and may reveal further possibilities for engineering the structure of the population by  varying the persistence length of individual constituent particles.

\acknowledgments
MJM acknowledges studentship funding from EPSRC through the Scottish CM-CDT under Grant No.~EP/L015110/1. For the purpose of open access, the author has applied a Creative Commons Attribution (CC BY) licence to any Author Accepted Manuscript version arising from this submission.


\end{document}